# Anharmonic Phonons and Anomalous Thermal Expansion of Graphite

Ranjan Mittal[1,2]*, Mayanak K. Gupta[1], Baltej Singh[1,2], S. K. Mishra[1] and Samrath L. Chaplot[1,2]
[1]*Solid State Physics Division, Bhabha Atomic Research Centre, Mumbai, 400085, India*
[2]*Homi Bhabha National Institute, Anushaktinagar, Mumbai 400094, India*
*Corresponding author, Tel: 0091 22 25594307, Email: rmittal@barc.gov.in (R. Mittal)*

We have investigated the anisotropic thermal expansion of graphite using ab-initio calculation of lattice dynamics and anharmonicity of the phonons, which reveal that the negative thermal expansion (NTE) in the a-b plane below 600 K and very large positive thermal expansion along the c-axis up to high temperatures arise due to various phonons polarized along the c-axis. While the NTE arises from the anharmonicity of transverse phonons over a broad energy range up to 60 meV, the large positive expansion along the c-axis occurs largely due to the longitudinal optic phonon modes around 16 meV and a large linear compressibility along the c-axis. The hugely anisotropic bonding in graphite is found to be responsible for wide difference in the energy range of the transverse and longitudinal phonon modes polarized along the c-axis, which are responsible for the anomalous thermal expansion behavior. This behaviour is in contrast to other nearly isotropic hexagonal structures like water-ice, which show anomalous thermal expansion in a small temperature range arising from a narrow energy range of phonons.

# 1. INTRODUCTION

Graphite has a layered structure with hexagonal symmetry [1]. The carbon atoms are arranged in each layer in a two-dimensional (2D) honeycomb lattice with extremely strong $sp^2$ intra-layer bonds. The layers interact through the van der Waals interactions. The highly anisotropic interaction in graphite leads to many novel physical and mechanical properties, such as maximal values of the electric and thermal conductivities, in-plane elastic stiffness and strength, and the minimum shear-to-tensile stiffness ratio. The understanding of the thermodynamic behaviour of graphite at the atomic level, as well as that of other graphitic materials, such as graphene and carbon nanotubes, is of fundamental interest [2-16]. The discovery of superconductivity on twisted layer graphene is also of considerable interest [17]. Spectroscopic studies[18-20] of the phonon spectrum have been reported using Raman, infrared and neutron scattering techniques. In situ measurement [21]of elastic and total strains during ambient and high temperature deformation of a polygranular graphite has been reported. The interfacial thermal conductance between graphite and substrate copper has been tuned[22] via anisotropic elastic properties of graphite. An understanding of lattice strain, defects and disorder in nuclear graphite has been reported [23]by x-ray diffraction and Raman measurements. Comparative study of melting of graphite and graphene has been performed [24] using both the empirical models and ab-initio simulations. Electronic structure of graphite is found to change[25] upon lithium intercalation. Ab-initio density-functional theory calculations of the structure, vibrational and thermodynamic properties of diamond, graphite and graphene have been performed[26] quite some time ago. However, only later theoretical and computational developments enabled a good description of the van der Waals dispersion forces by the exchange correlation functionals used in the density functional theory formalism.

Graphite is known to show anomalous thermal-expansion behavior [27-30], where thermal expansion coefficient is negative in the a-b plane and positive along the c-axis. To address this anomalous behavior in graphite, we have performed extensive ab-initio phonon calculations and explained the role of anharmonic phonons leading to the anomalous behaviour. As noted above, the van der Waals interactions [13] are very important for a proper description of the graphite structure and its properties, and these have been suitably included in our ab-initio calculations.

# 2. COMPUTATIONAL DETAILS

The phonon frequencies in the entire Brillouin zone are obtained using the PHONON5.2[31] software package in conjunction with the Vienna-based ab-initio simulation package (VASP)[32, 33]. A supercell approach has been used to calculate the phonon force constants between various atomic pairs, which is further used to calculate the dynamical matrix and hence the phonons in entire Brillouin zone. The PHONON5.2[31] software package has been used to generate the supercell with various atomic displaced configuration. A supercell of 8×8×2 has been used for the phonon calculation. The force calculation of distinct configuration are performed using the Vienna based ab-initio simulation package[32, 33] (VASP). The exchange correlation part of DFT formalism is treated with generalized gradient approximation with Perdew, Burke and Ernzerhof[34, 35] (PBE) functional. Further to incorporate the interaction between the valance and core electrons the projected augmented wave method is used. A plane wave basis set cutoff of 1000 eV was used for basis set expansion. The Brillouin zone integrations were performed on 20×20×2 mesh generated using the Monkhorst-Pack method[36]. The electronic convergence and ionic forces tolerance were chosen to be $10^{-8}$ eV and $10^{-3}$ eV/Å respectively. All the calculations are performed with relaxed geometry. The relaxation of the graphite structure was performed with including various vdW-DFT non-local correlation functional[37-39]. We find that the relaxation performed using the optB88-vdW[39] (BO) functional scheme of the vdW-DFT method produces the best match with the experimental structure.

We have also performed ab-initio molecular dynamics (AIMD) simulation at 300 K and 1000K on 4×4×1 cell (64 atoms) using NVT ensemble. For AIMD calculations, gamma point k-mesh with an electronic convergence criterial of $10^{-6}$ eV were used. The non-local part of the pseudopotential was computed with real-space projection scheme in order to accelerate the AIMD simulation. A 2 fs timestep were chosen to integrate the equation of motion. We performed the simulation up to 15 pico-second for 300 K and 1000 K, where the first 3 picosecond trajectories were discarded for equilibration purpose. To control the temperature of the simulation, we employed the Nose Hover thermostat[40] with Nose mass 0.16 amu. The time dependence of the trajectories and forces were used to extract the renormalized force constants using ALAMODE[41]. We have used 8.5 Å and 6.0 Å for the second and third order force constant cutoff, respectively. These renormalized force constants were used to compute the phonon dispersion and density of states of graphite at 300K and 1000K.

## 3. RESULTS AND DISCUSSION

### 3.1 Ab-initio Calculation of the Phonon Spectrum

The calculated values of the a- and c-lattice parameter as obtained from structural relaxation of graphite are 2.465 Å and 6.693 Å respectively, which match very well with the experimental values [1] of 2.462 Å and 6.707 Å respectively [1]. The phonon modes along various high symmetry directions have been measured by neutron[42] as well as x-ray inelastic scattering [43, 44]. The calculated dispersion relation (**Fig. 1**) is in very good agreement with the available experimental data. The calculated phonon density of states for graphite is shown in **Fig. 2**. The calculated spectra have peaks at 58 meV, 78 meV, 105 meV and 170 meV. The peaks in the phonon density of states as well as the energy-range of the phonon spectrum are in very good agreement with the experimental data[19] from neutron inelastic scattering. We have also calculated the separate contributions to the phonon spectrum from the in-plane and out-of-plane vibrations. The results show that the contribution below 110 meV is largely from the out-of-plane vibrations of the carbon atoms. The strong peak in the spectrum at about 180 meV is due to the in-plane vibrations.

### 3.2 Ab-initio Calculations of Thermal Expansion Behavior

Graphite shows anomalous thermal-expansion behavior [27-30, 45-47]. The thermal expansion coefficient is found to show small negative values in the a-b plane below 600 K and large positive values along the c-axis. Theoretical models [27] have been used to understand the thermal expansion behavior of graphite. However, still there is a lack of microscopic understanding of the phonon modes which govern the anomalous behavior. We use ab-initio calculations to understand the thermal expansion behavior of graphite. The pressure dependence of phonon frequencies in the entire Brillouin zone[48] in the quasi-harmonic approximation has been used for the calculation of the linear thermal expansion coefficients, which are given by the following relation[49] for a hexagonal system:

$$\alpha_l(T) = \frac{1}{V_0} \sum_{q,j} C_v(q,j,T) \left[ (s_{l1} + s_{l2})\Gamma_a + s_{l3}\Gamma_c \right] , \quad l, l' = a(= b), c \ \& \ l \neq l' \quad (1)$$

$s_{ij}(=C_{ij}^{-1})$ are elements of the elastic compliances matrix at constant temperature T=0 K, $C_{ij}$ are the elastic constants, $V_0$ is volume at 0 K and $C_v(q, j, T)$ is the specific heat at constant volume due to $j^{th}$ phonon mode at point **q** in the Brillouin zone. The anisotropic mode Grüneisen parameters of the phonon of energy $E_{q,j}$ are given as:

$$\Gamma_l(E_{q,j}) = -\left(\frac{\partial ln E_{q,j}}{\partial ln l}\right)_{l'}; \; l, l' = a(=b), c \; \& \; l \neq l' \qquad (2)$$

For a hexagonal system, $\Gamma_a = \Gamma_b$. The volume thermal expansion coefficient for a hexagonal system is given by:

$$\alpha_V = (2\alpha_a + \alpha_c) \qquad (3)$$

The anisotropic stress dependence of the energies of phonon modes has been calculated to estimate the anisotropic mode Grüneisen parameters ($\Gamma_a$, $\Gamma_c$). **Figure 3(a)** gives a plot of the values of $\Gamma_a$ and $\Gamma_c$ as a function of energy averaged over phonons of the same energy in the entire Brillouin zone. The mode Grüneisen parameters $\Gamma_a$ shows a large variation in its magnitude. For phonons of energy below 5 meV, $\Gamma_a$ has very large positive values. For higher energy phonons from 10 to 25 meV, $\Gamma_a$ varies from -10 to +5. In contrast to this, $\Gamma_c$ shows smaller variation with the maximum value of +15 for low energy phonons below 10 meV.

In **Table I**. we have presented the calculated elastic constants and the elastic compliance matrix elements, and also the measured elastic constants [50] for comparison with the calculations. The large anisotropy bonding between various carbon atoms within the *a-b* plane and along c-axis results in large anisotropic values of Cij, which is in agreement with the experimental values.

The calculated linear thermal expansion coefficients $\alpha_l(T)$ are shown in **Fig. 3(b)**. The large anisotropic behaviour of $\alpha_l(T)$ results from a combination of large anisotropy in the calculated elements of elastic compliances and the anisotropic mode Grüneisen parameters ($\Gamma_a$, $\Gamma_c$) according to Eq. (1). The calculations reproduce fairly well (**Fig. 3(b)**) the experimental values of $\alpha_a$, while $\alpha_c$ values are found to be underestimated in comparison to the experimental values. We note that the quasi-harmonic approximation accounts for the volume-dependent anharmonicity but not for the explicit temperature-dependent anharmonicity. We have shown below in Section 3.3 that there is indeed significant variation of the phonon energies as a function of temperature. In a previous calculation [26], it is reported that the bulk modulus changes by as much as 25 % from a value of 39.5 GPa at 0 K to 30 GPa at 1000 K. The underestimation in the calculated thermal expansion at high temperatures is expected in materials that show a significant explicit anharmonicity.

In order to understand the role of anisotropy of the elastic behavior and the mode Grüneisen parameters on the linear thermal expansion coefficients, we can re-express Eq. (1) using the elastic compliance values (TABLE I) as follows:

$$\alpha_a(T) = \frac{1}{V_0}\Sigma_{q,j} C_v(q,j,T)\,[0.80\,\Gamma_a + 0.10\Gamma_c] \quad (4)$$

$$\alpha_c(T) = \frac{1}{V_0}\Sigma_{q,j} C_v(q,j,T)\,[0.20\,\Gamma_a + 24.30\,\Gamma_c] \quad (5)$$

It can be seen that large anisotropy in elastic compliances matrix (TABLE I) results in important role of first and second terms in the above equations in governing the sign of $\alpha_a(T)$ and $\alpha_c(T)$ respectively. $\Gamma_a$ has very large positive values (**Fig. 3(a)**) for phonons of energy below 5 meV. Although these modes have large positive $\Gamma_a$ values, their contribution to the total phonon spectrum is very low, which results in low positive $\alpha_a(T)$ values below 70 K. At higher temperatures, the contribution from high energy modes with negative $\Gamma_a$ becomes important and results in negative thermal expansion behaviour up to 440 K. We find that the matrix element $S_{33}$ has a very large value of in comparison to other matrix elements, which leads to large linear compressibility along the c-axis. This results in large positive values of $\alpha_c(T)$, although the magnitude of positive values for $\Gamma_c$ is relatively small.

In order to identify the phonons that are most relevant to the thermal expansion, we have calculated the contribution to the volume thermal expansion coefficient from phonons of energy E averaged over the entire Brillouin zone as a function of phonon energy (E) at 200 K (**Fig. 4**). The phonons of energy around 16 meV are found to contribute maximum to $\alpha_c$, while those over a broad range of energies up to about 60 meV contribute to $\alpha_a$. As discussed above, the low-energy acoustic phonons below 5 meV do not have much weight in the total spectrum, and so their contribution is small despite their large positive $\Gamma_a$.

As discussed above, the phonons in the entire Brillouin zone contribute[48, 51, 52] to the total thermal expansion behavior and its nature (positive or negative) is governed by dominating contribution of various phonon modes. In **Fig. 5**, we plot the Grüneisen parameters $\Gamma_a$ and $\Gamma_c$ for phonons along the high-symmetry directions. The large negative values of $\Gamma_a$ occur for the transverse acoustic (R1) and the transverse optic (R2) branch, both of them polarized along the c-axis. These branches extend up to about 60 meV and contribute to the NTE in the a-b plane. The same branches also have large positive $\Gamma_c$ values and contribute to the positive thermal expansion along the c-axis. However, a large positive contribution to $\Gamma_c$ is made by the phonon branches propagating along the c-axis around 16 meV, which include the longitudinal acoustic and the longitudinal optic branch.

Several sharp peaks in the phonon density of states are observed at different energies; these well separated peaks arise from the anisotropic $sp^2$ interaction behavior of carbon atoms in graphite. Several other hexagonal structured compounds are also known to exhibit anomalous thermal expansion behavior. In our previous studies, we investigated interesting lattice dynamics behavior of variety of hexagonal structured compounds and unraveled the role of phonon anharmonicity in the thermodynamic behaviour on those compounds. A notable compound is water-ice that also crystallizes in hexagonal structure below 273 K and known as ice *Ih* (space group P6$_3$/mmc). It exhibits NTE [53, 54] below ~ 60 K. We discovered that [55] phonons of energy around 6 meV are the key contributer to the observed NTE behaviour in ice *Ih*. These phonon modes can be visualized as rotational motion of the hexagonal rings of ice molecules and the transverse vibrations of the hexagonal layers, and mainly contribute to the NTE in ice *Ih*.

Another very interesting hexagonal structured compound is ZnAu$_2$(CN)$_4$ (space group P6$_2$22), which shows [56, 57] anisotropic thermal expansion behaviour ($\alpha_a$~ 36.9×10$^{-6}$ K$^{-1}$, $\alpha_c$~ -57.6×10$^{-6}$ K$^{-1}$) as well as unusually large negative linear compressibility. Here also, we found that [58] the phonon modes of energy around 5±1 meV contribute maximum to the anisotropic positive and negative thermal expansion behavior. However, their dynamics is different from that in ice. These modes involve perpendicular displacements of Au, C and N atoms to the Zn(CN)$_4$-Au-(CN)$_4$Zn linkage. The difference in the dynamical character of the low-energy modes in ZnAu$_2$(CN)$_4$ from that in ice *Ih* is mainly attributed to the difference in the atomic arrangement/structure in these compounds. The compound ZnAu$_2$(CN)$_4$, is formed by double helical chain of Zn(CN)$_4$ tetrahedra interconnected by dicyanoaurate (NC-Au-CN) linkage, in contrast to the hydrogen-bonded structure of ice *Ih*. Hence the difference in the structure and the underlying anisotropic behavior of the interatomic interactions leads to the different vibrational pattern leading to anomalous properties.

As mentioned above the bonding in graphite is quite different from other hexagonal structured compounds. The strong bonding in the a-b plane results in the transverse acoustic modes to have energies up to high-energy of 60 meV and their contribution governs the NTE behavior of graphite.

We have investigated the displacement pattern (eigenvectors) of a few representative low-energy phonon modes which have dominating contribution at low temperature ~ 200 K to the anomalous thermal expansion in graphite. We have identified a few representative modes at (0.25 0 0), Γ-point (0 0 0) and A-point (0 0 0.5) of energies 19.8 meV, 16.6 meV and 11.9 meV, respectively, which contribute to NTE along the a-axis and expansion along the c-axis. The contribution to the linear thermal expansion coefficients at 200 K from each of these modes is given below, assuming each of the modes as an Einstein mode with one degree of freedom:

(0.25 0 0), 19.8 meV:  $\alpha_a$= -3.8 ×$10^{-6}$ $K^{-1}$, $\alpha_c$=5.6×$10^{-6}$ $K^{-1}$
(0 0 0), 16.6 meV:     $\alpha_a$= -0.1 ×$10^{-6}$ $K^{-1}$, $\alpha_c$=52.3×$10^{-6}$ $K^{-1}$
(0 0 0.5), 11.9 meV:   $\alpha_a$= -0.1 ×$10^{-6}$ $K^{-1}$, $\alpha_c$= 55.0 ×$10^{-6}$ $K^{-1}$

It can be seen that the displacements of atoms along the c-axis (**Fig.6**) in the mode at (0.25 0 0), and equivalently, at (0 0.25 0) produce a shear wave in the a-b plane, which results in compression of the lattice in the a-b plane and expansion along the c-axis. The Γ-point and A-point modes involve vibrations of carbon atoms along the c-axis, which cause large thermal expansion along the c-axis.

### 3.3 Temperature Dependence of Phonon Spectrum

Temperature and pressure dependence of the energy of phonon modes is known to occur due to anharmonicity of the interatomic potential. The "implicit" anharmonicity is related to the volume dependence of the phonon spectrum, while the "explicit" anharmonicity arises due to large thermal amplitude of atoms. Temperature-dependent measurements involve both the effects [48] due to the implicit as well as explicit anharmonicities.

In order to understand the role of the explicit and implicit anharmonicities on the phonon energies on increase of temperature we have performed AIMD calculations for graphite at 300 K and 1000 K. We have used the method of Ref. [41], where these effects are considered harmonically and give rise to renormalized phonon frequencies. The shift in the phonon frequencies at finite temperature with respect to 0 K is associated with higher order anharmonicity. The calculations at both the temperatures are performed using the structure parameters (a= 2.465 Å and c=6.693 Å) as used in ab-initio lattice dynamics calculation of the phonon spectrum of graphite. Since volume in both the calculations is the same, the changes in the phonon spectrum would be only due to the explicit anharmonicity due to large thermal amplitude of atoms.

The calculated temperature-dependent phonon dispersion relation along the high symmetry directions of graphite at 300 K and 1000 K are shown in **Fig.7**. The calculated dispersion relation at 300 K (**Fig.7**) only qualitatively matches with the phonon dispersion calculated from the ab-initio lattice dynamics (**Fig. 1**). The renormalized phonons at 300 K and 1000 K show imaginary transverse acoustic modes near the zone center along (001) direction, which is due to numerical errors associated with AIMD forces using a single K-point integration and the real-space projection scheme. Since the low-energy modes are associated with inter-planer forces of carbon atoms, which are much smaller in comparison to the intra-planer carbon forces in graphite, hence even small errors in the force calculation may lead to significant deviation in inter-planer force constants. However, our calculations are useful for understanding, at least qualitatively, the role of temperature in leading to change of phonon energies in graphite. The comparison of the calculations at 300 K and 1000 K clearly reveals significant explicit anharmonicity of most phonons. The calculated phonon density of states (**Fig. 8**) at 300 K and 1000 K shows that peaks in the density of states at 60 meV, 110 meV and 170 meV are found to shift to lower energies with increase of temperature due to the explicit anharmonicity. The quasi-harmonic lattice dynamics calculations, as discussed above, are found to underestimate the thermal expansion at high temperatures, since these do not account for the explicit anharmonicity of phonons. The ab-initio molecular dynamics calculation confirms that the explicit anharmonicity is indeed significant in graphite at elevated temperatures.

## 4. CONCLUSIONS

Ab-initio calculations are performed to understand the origin of anomalous thermal expansion behavior in graphite. The calculations are able to reproduce the highly anisotropic thermal expansion involving a small negative expansion in the a-b plane and very large thermal expansion along the c-axis. This behaviour may be attributed to anharmonic vibrations of atoms along the c-axis, besides a large linear compressibility along the c-axis. While the NTE is found to be mainly due to the transverse-acoustic and transverse-optic modes polarized along the c-axis over 0-60 meV, the large expansion along the c-axis occurs mainly due to the longitudinal-optic modes polarized along the c-axis that occur over a narrow-range of energies around 16 meV. The large difference in the range of energies of the transverse and the longitudinal phonon modes responsible for the anomalous thermal expansion behavior is found to be due to hugely anisotropic bonding in graphite, namely, the covalent bonding in the a-b plane and the van der Waals bonding along the c-axis. Such a behaviour is in sharp contrast to that in other nearly isotropic-bonded hexagonal materials like water-ice. The quasi-harmonic approximation (QHA) used in the lattice dynamics calculations is found to underestimate the thermal expansion at high temperatures.

This is often the case since QHA includes only the implicit anharmonicity due to the change in volume, but it does not include the effect of explicit anharmonicity due to the increase in thermal amplitudes. Using ab-initio molecular dynamics as a function of temperature, but at a fixed volume, we find that there is indeed a large explicit anharmonicity of phonons, which may be the reason for the underestimate in the calculated thermal expansion at high temperatures.

**Notes**

There are no conflicts to declare

**CRediT authorship contribution statement**

All the authors contributed in formulation of the problem, planned, performed ab-initio calculations, interpreted various results of ab-initio calculations, and contributed in writing of manuscript.

**Declaration of competing interest**

The authors declare that they have no known competing financial interests or personal relationships that could have appeared to influence the work reported in this paper.

**Acknowledgements**

The use of ANUPAM super-computing facility at BARC is acknowledged. SLC thanks the Indian National Science Academy for award of an INSA Senior Scientist position.11

TABLE I. The calculated and experimental [50] elastic constants ($C_{ij}$) of graphite in GPa units, and the elements of elastic compliances matrix ($s = C^{-1}$) in the units of $10^{-3}$ GPa$^{-1}$.

| Indices ij | Experimental $C_{ij}$ | Calculated ab-initio | |
|---|---|---|---|
| | | $C_{ij}$ | $s_{ij}$ |
| 11 | 1060 | 1058.6 | 1.0 |
| 33 | 36.5 | 41.1 | 24.3 |
| 12 | 180 | 191.8 | -0.2 |
| 13 | 15 | -3.5 | 0.1 |
| 44 | 4 | 4.5 | 222.8 |
| 66 | 440 | 433.2 | 2.3 |

**Fig. 1** The calculated phonon dispersion relation of graphite. The phonon modes (solid circles) are measured by neutron [42] as well as inelastic x-ray scattering [43, 44].

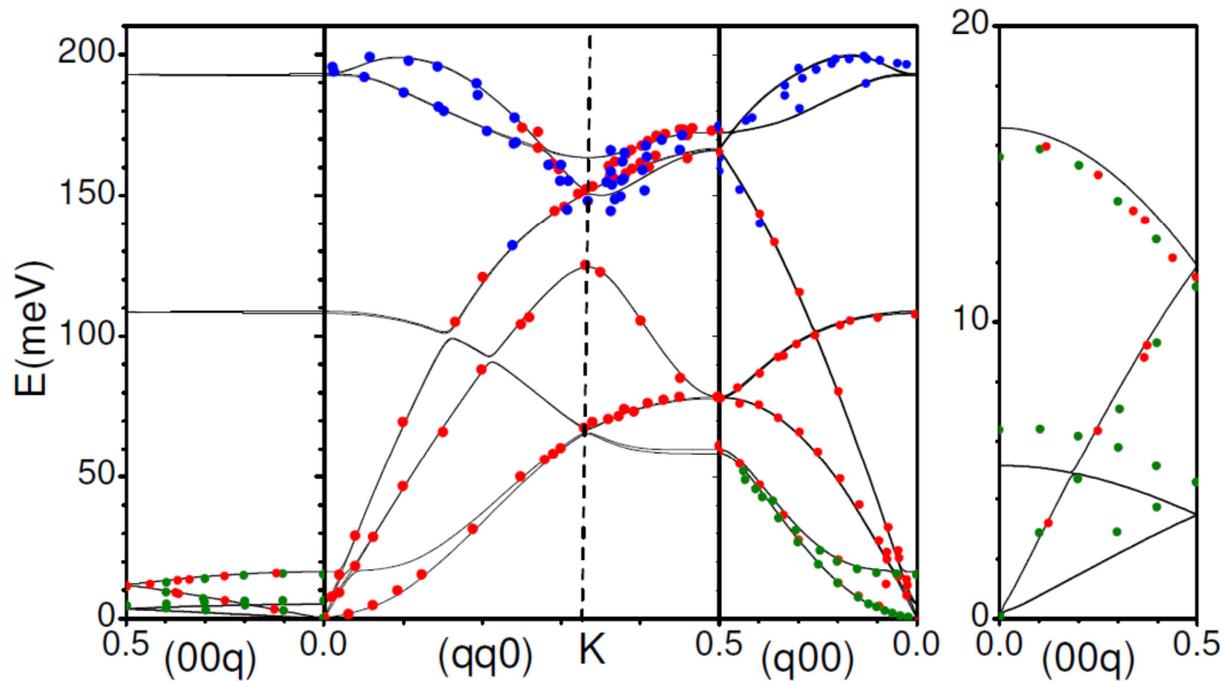

**Fig. 2** The calculated phonon density of states of graphite and the partial phonon density of states. $g_x$ and $g_z$ are the components of the partial phonon density of states in the a-b plane and the c-axis respectively. The experimental data of phonon density of states for graphite are from Ref. [19]. The calculated spectra are broadened by a Gaussian of FWHM of 6 meV.

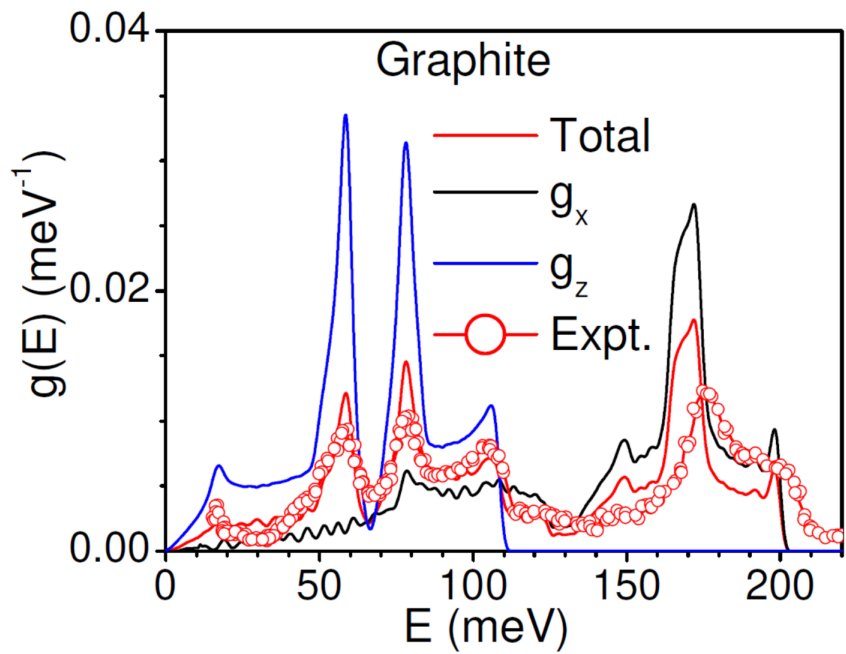

**Fig. 3** (a) The calculated anisotropic mode Grüneisen parameters $\Gamma_a$ and $\Gamma_c$ of phonons of energy E averaged over the entire Brillouin zone as a function of phonon energy (E) in graphite. (b) The calculated linear thermal expansion coefficients $\alpha_l$ ($l$=a,c) in graphite. The available experimental data [28, 45-47] (open and closed symbols) from the literature in graphite and pyrolytic graphite are also shown.

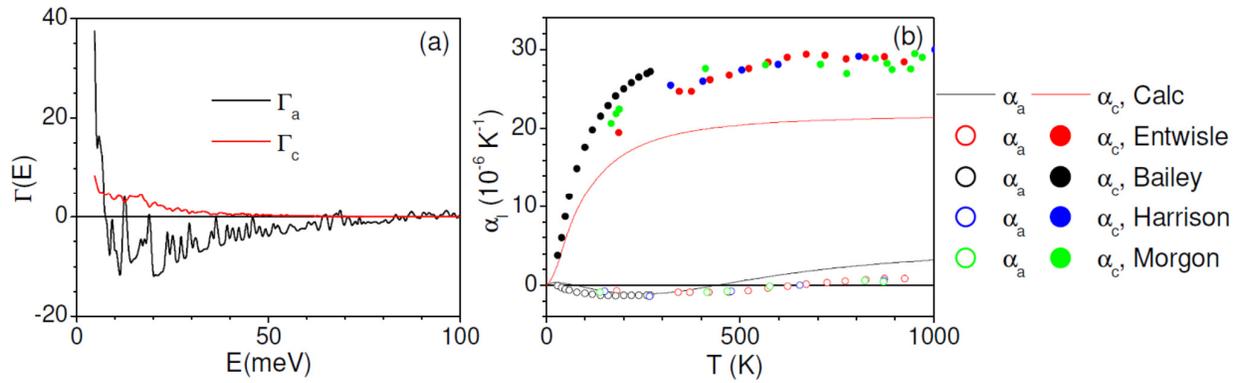

**Fig. 4** The calculated contribution to the linear thermal expansion coefficients from phonons of energy E averaged over the entire Brillouin zone as a function of phonon energy (E) at 200 K in graphite.

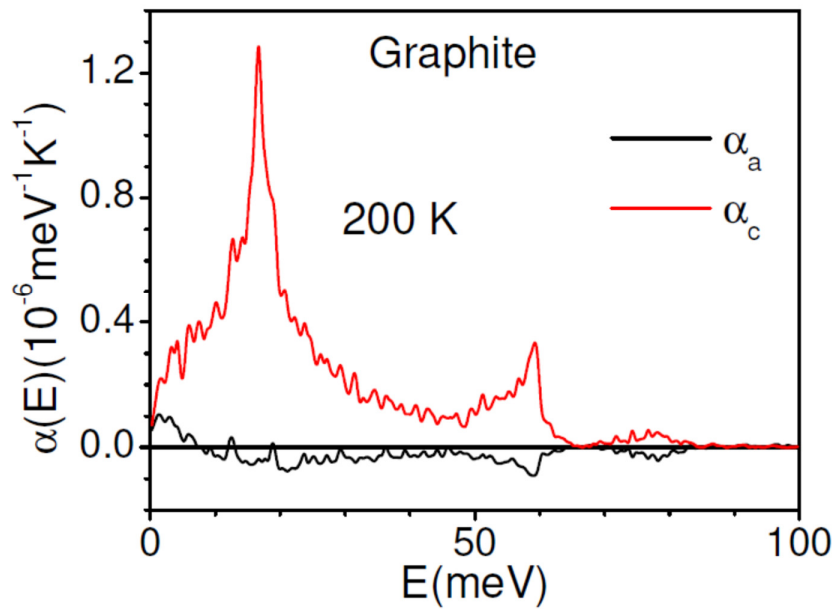

**Fig. 5** The calculated anisotropic Grüneisen parameters ($\Gamma_a$, $\Gamma_c$) for phonon branches up to 120 meV along high symmetry directions. The calculated phonon dispersion branches are shown by various colors. The colors' scheme shown in the calculated $\Gamma_a$, $\Gamma_c$ plots corresponds to the colors used to identify the phonon branches in the phonon dispersion relation.

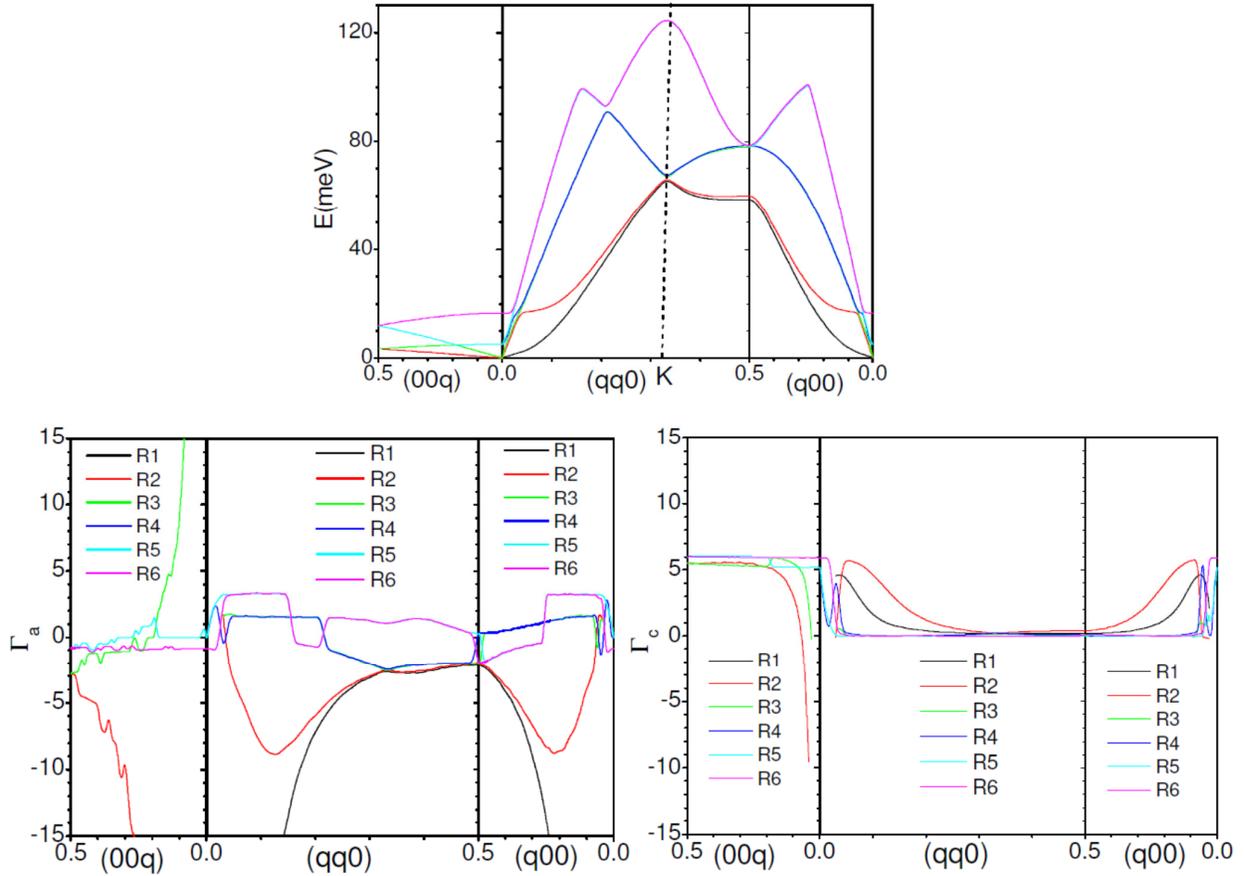

**Fig. 6** The atomic displacement patterns of selected phonon modes in graphite. The numbers below the figures give the phonon wave vector, phonon energy, $\Gamma_a$, $\Gamma_c$ and contribution of the specific phonon as an Einstein oscillator to $\alpha_a$ and $\alpha_c$ respectively. The values of linear thermal expansion coefficients ($\alpha_a$ and $\alpha_c$) are at 200 K and are in the units of $10^{-6}$ K$^{-1}$.

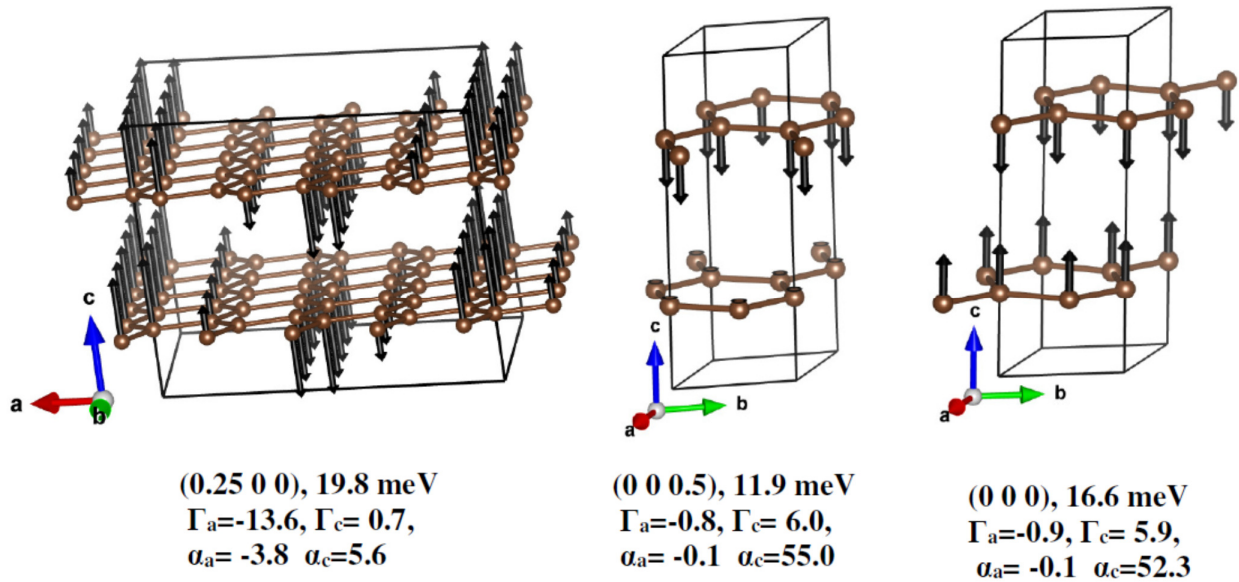

(0.25 0 0), 19.8 meV
$\Gamma_a$=-13.6, $\Gamma_c$= 0.7,
$\alpha_a$= -3.8  $\alpha_c$=5.6

(0 0 0.5), 11.9 meV
$\Gamma_a$=-0.8, $\Gamma_c$= 6.0,
$\alpha_a$= -0.1  $\alpha_c$=55.0

(0 0 0), 16.6 meV
$\Gamma_a$=-0.9, $\Gamma_c$= 5.9,
$\alpha_a$= -0.1  $\alpha_c$=52.3

**Fig. 7** The calculated temperature dependence of the phonon dispersion relation of graphite from ab-initio molecular dynamics simulations at a fixed volume, indicating the explicit anharmonicity due to increase in thermal amplitudes only.

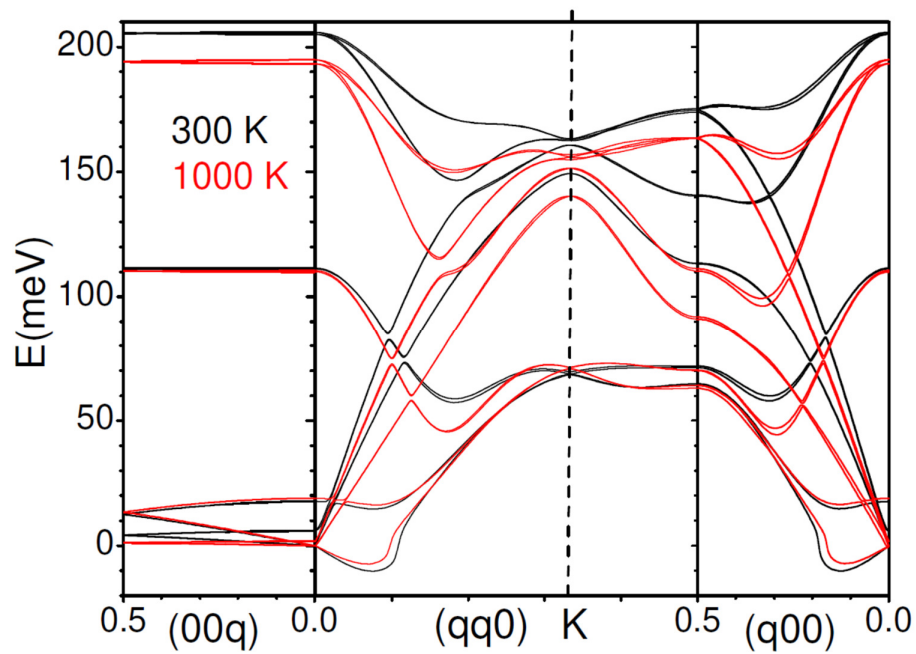

**Fig. 8** The calculated temperature dependence of the phonon density of states of graphite from ab-initio molecular dynamics simulations at a fixed volume, indicating the explicit anharmonicity due to increase in thermal amplitudes only.

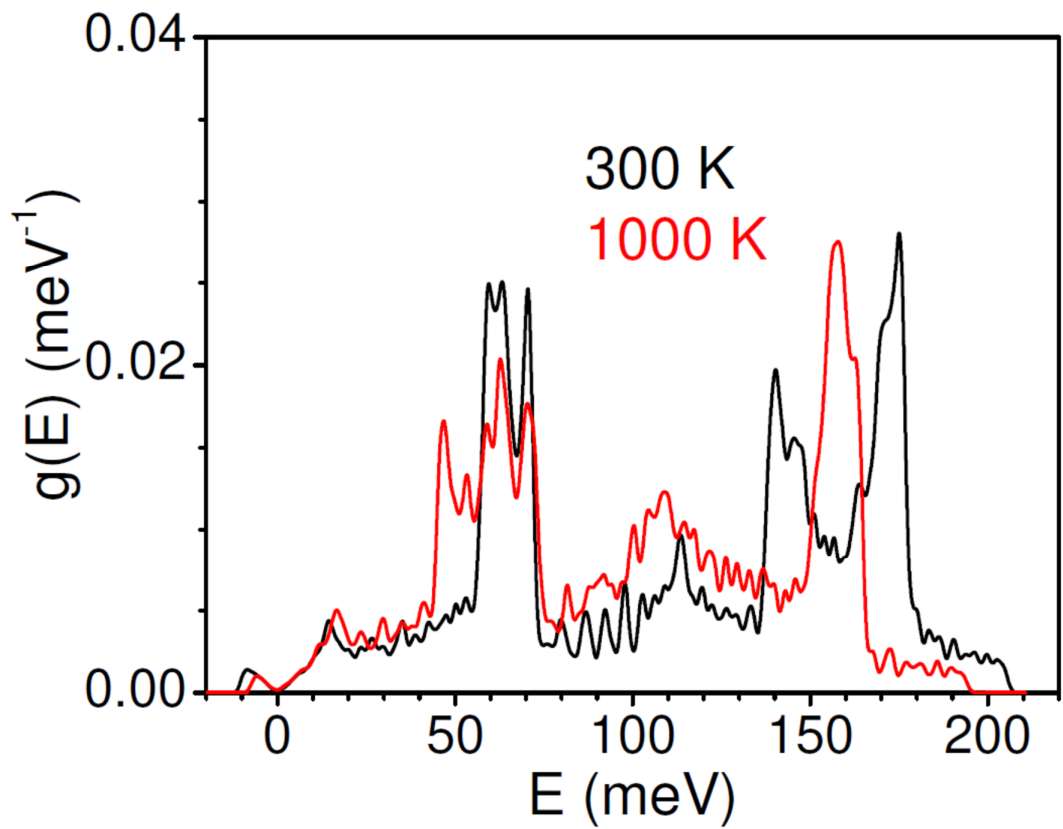